# A study of co-movements between oil price, stock index and exchange rate under a cross-bicorrelation perspective: the case of Mexico

Semei Coronado[1], Omar Rojas[2]

## Abstract

In this chapter we studied the nonlinear co-movements between the Mexican Crude Oil price, the Mexican Stock Market Index and the USD/MXN Exchange Rate, for the sample period from 1994 to date. We used a battery of nonlinear tests, cf. (Patterson & Ashley, 2000) and one multivariate test, in order to determine the dynamic co-movement exerted from the oil prices to the stock and exchange rate markets. Such co-movement and time windows are exposed using the Brooks & Hinich (1999) cross-bicorrelation statistical test. The effects of oil spills on other markets have been studied from different angles and on several financial assets. In this study, we focus our attention on the detection, not only of the correlations amongst markets but on the epochs in which such nonlinear dependence might occur. This is important in order to understand better, how the markets that drive the economy interact with each other. We hope to contribute to the literature with such findings, filling a gap in the emerging markets context, in particular, for the Mexican case.

[1] Universidad de Guadalajara, Profesor del Departamento de Métodos Cuantitativos, México; e-mail: semeic@cucea.udg.mx
[2] Universidad Panamericana, Campus Guadalajara, Profesor-Investigador de la Escuela de Ciencias Económicas y Empresariales, México; e-mail: orojas@up.edu.mx

# 1. Introduction

The study of oil prices is of vital importance, given the high volatility levels that it has reached in the last years and the impact it has on other financial markets and macroeconomic variables of all countries of the world and, in particular, in importing and exporting countries, as is the case of Mexico. Furthermore, oil is a product from which other products are derived, so they can be used in different ways, having an impact on financial markets and exchange rates; the fluctuation of oil prices might affect the global economy behavior: raising the costs of production of goods and services, having and impact on inflation, transfer of wealth from oil consumers to oil products and consumer's trust, amongst others, cf. (Blanchard & Gali, 2007; Jiménez-Rodríguez & Sánchez, 2005; Turhan, Hacihasanoglu, & Soytas, 2013).

If oil prices have a spillover impact on financial markets, then it would be expected for such prices to be correlated to the prices of stocks. As Jones *et al.* (2004) states, "Ideally, stock values reflect the market's best estimate of the future profitability of firms, so the effect of oil price shocks on the stock market is a meaningful and useful measure of their economic impact. Since asset prices are the present discounted value of the future net earnings of firms, both the current and the expected future impacts of an oil price shock should be absorbed fairly quickly into stock prices and returns, without having to wait for those impacts to actually occur." Therefore, if there is a rise on the oil prices, the stock, markets would experience a decrease on their returns, depending on whether it is an importing or exporting country: for an importing country, the rise on the oil price puts decreasing pressure on the exchange rate and a rise on the internal inflation (R. D. Huang, Masulis, & Stoll, 1996); in this way, oil prices are related to exchange rates due to import and exports of consumers and producers, as already pointed out by Golub (1983) and Krugman (1983), and more recent studies, cf. (Akram, 2004; Amano & Van Norden, 1998; Bénassy-Quéré, Mignon, & Penot, 2007; Lizardo & Mollick, 2010; Park & Ratti, 2008; Reboredo, Rivera-Castro, & Zebende, 2014). Therefore, the study of the relationship between oil, exchange rate and financial markets, in particular in the case of Mexico, if of great interest, given that Mexico is an oil producer and considered a growing emerging economy.

There are different studies on the relationships between oil prices and macroeconomic variables, specifically between exchange rate and financial markets. Such studies have been done in developed and emerging markets, and in countries that import and export oil, view from different economic and financial perspectives, if the markets are efficient. Regime switching thresholds, supply and demand equilibrium points, long-run relationships, monetary and fiscal policy impacts on oil prices, appreciation or depreciation of currency usually with respect to US dollar or Euro provide better understanding of portfolios for stock holders and managers, amongst others, cf. (Abdalla & Murinde, 1997; Syed A Basher & Sadorsky, 2006; Syed Abul Basher, Haug, & Sadorsky, 2012; Eryigit, 2012; B.-N. Huang, Hwang, & Peng, 2005; Imarhiagbe, 2010; Jawadi & Leoni, 2009; Liu, Ji, & Fan, 2013; Narayan & Narayan, 2010; Sahu, Mondal, & others, 2015).

In the case of Mexico, there are some studies that have focused their attention on the relationship of the oil price and the financial market, applying different methodologies, cf. (Filis, Degiannakis, & Floros, 2011; Hammoudeh & Li, 2005; Maghyereh, 2004; Mohanty, Nandha, Turkistani, & Alaitani, 2011); for the relationship between oil price and exchange rate cf. (Turhan et al., 2013), whereas for the relationship between exchange rate and financial market, cf. (Diamandis & Drakos, 2011; Ocampo, 2009; Volkov & Yuhn, 2015).

The research presented in this chapter analyzed the case of Mexico, given that it is an oil exporting country and a growing emerging economy. In this chapter we study the nonlinear co-movements between the Mexican Crude Oil Price –not the Brent or the West Texas International–, the Mexican Stock Market Index (IPC), and the USD/MXN Exchange Rate, for the sample period from 1994 to date. We used a battery of nonlinear tests, cf. (Patterson & Ashley, 2000) and one multivariate test, in order to determine the dynamic co-movement exerted from the oil prices to the stock and exchange rate markets. Such co-movement and time windows are exposed using the Brooks & Hinich (1999) cross-bicorrelation statistical test, which is a non-linear multivariate test that uses cross-bicorrelation, in order to determine the functional non-lineal relationship between a pair of time series. It uncovers the direction of flow for such bi-correlations and the significant lag through the third order distribution moments. This test has been successfully applied to different economic and financial nonlinear time series, cf.

(Coronado-Ramírez, Rojas, Romero-Meza, & Venegas-Martinez, 2015; Czamanski, Dormaar, Hinich, & Serletis, 2007; Romero-Meza, Coronado, & Serletis, 2014; Serletis, Malliaris, Hinich, & Gogas, 2012).

The effects of oil spills on other markets have been studied from different angles and on several financial assets. In this study we focus our attention on the detection, not only of the correlations amongst markets, but on the episodes in which such nonlinear dependence might occur. This is important in order to understand better how the markets that drive the economy interact with each other. We hope to contribute to the literature with such findings, filling a gap in the emerging markets context, in particular for the Mexican case.

The structure of the document is as follows: in section 2 we study the time series of prices and returns from a univariate point of view, showing descriptive statistics, stationarity and nonlinearity tests; in section 3 we describe the methodology used, namely the Brooks and Hinich cross-bicorrelation test; section 4 presents the results and discussion of the empirical analysis, whereas section 5 concludes.

2. **Empirical Data**

Three time series prices are analyzed in this paper: Mexican Mayan Crude Oil (Oil), Exchange Rate US dollar/ Mexican peso (USD/MXN) and Mexican Stock Market Index (IPC). All prices were collected from *Bloomberg*. The time period analyzed spans from February 01, 1998 to January 30, 2014, for a total 4,278 observations per series.

In order to make the data usable for the analysis, they were transformed to eliminate any linear dependencies. The resulting transformed series were used to test for co-movements and dependency. Next we show the data transformation process.

As a first step, prices were transformed into a series of continuously compounded percentage returns, by taking the difference of the logarithms of consecutive observations, *i.e.* $r_t = 100 \, (\ln(p_t) - \ln(p_{t-1}))$, where $p_t$ is the price of the series on day $t$. Fig. 1 shows the prices and returns for all series.

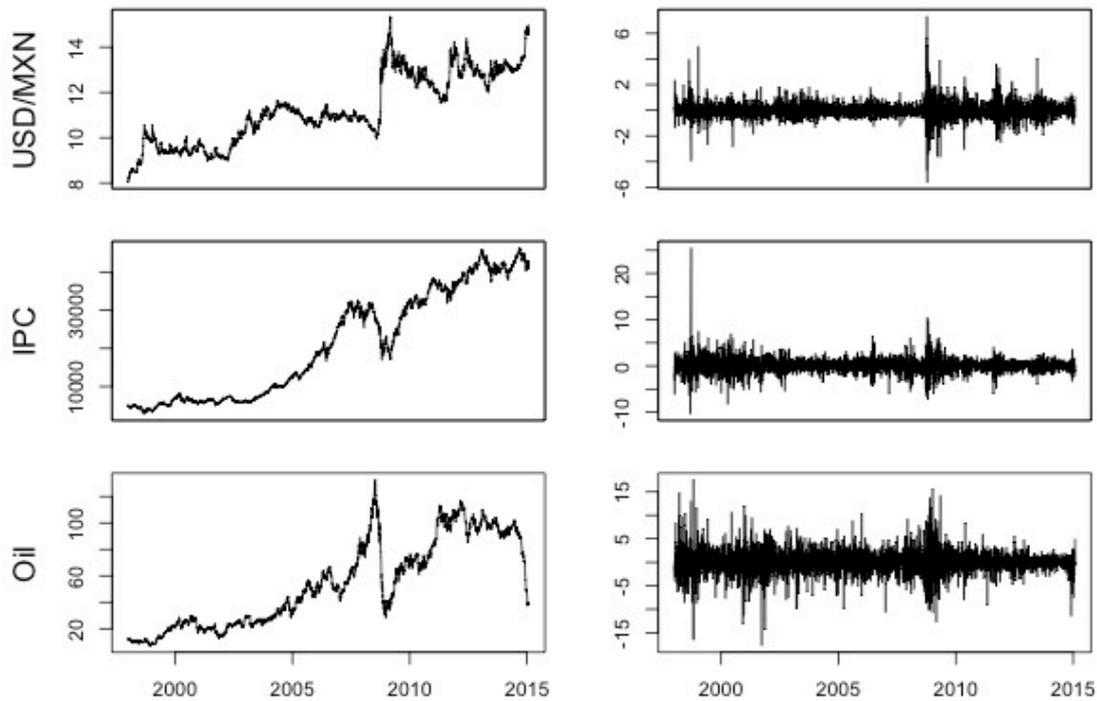

**Fig. 1** Time series plot of prices (left) and returns (right) for USD/MXN (top), IPC (middle) and Oil (bottom). Source: Authors elaboration using *Bloomberg* data.

## 2.1. Descriptive Statistics

Summary statistics for the returns are shown in Table 1. The statistics are consistent, as expected, with some of the stylized facts of financial time series (Cont, 2001). In particular, the kurtosis indicates that return distributions are leptokurtic. Furthermore, the Jarque-Bera statistic confirms returns not normally distributed.

**Table 1.** Summary statistics for the return time series

| Statistics | Oil returns | USD/MXN returns | IPC returns |
|---|---|---|---|
| Observations | 4277 | 4277 | 4277 |
| Mean | 0.03 | 0.01 | 0.05 |
| Standard Deviation | 2.45 | 0.63 | 1.51 |
| Skewness | -0.04 | 0.77 | 1.10 |
| Kurtosis | 6.09 | 13.91 | 22.55 |
| Jarque-Bera | 1699.89 | 21639.46 | 68938.14 |

## 2.2. Stationarity tests

In table 2, we present results of the ADF (Dickey & Fuller, 1981) and RALS (Im, Lee, & Tieslau, 2014) tests. As it is well known, as more negative the value of the statistic, the stronger the rejection of the hypothesis, that there is a unit root at the 1%, 5% and 10% significance level chosen. As we can observe, both tests imply rejection of the null hypothesis.

Table 2. Results from the ADF and RALS stationarity tests.

| Series | Statistic | |
|---|---|---|
| | ADF | RALS |
| Oil | -14.88 | -15.82 |
| USD/MXN | -17.73 | -19.09 |
| IPC | -17.29 | -18.37 |

*significant at 1%, 5% and 10%. The critical value for the ADF test is -3.96,-3.41,-3.12 respectively

## 2.3. Tests of nonlinear serial dependence

We present the results of the three nonlinear tests applied to the sample, from the battery of nonlinear tests of Patterson and Ashley (2000): McLeod and Li (1983), Engle LM (1982) and BDS (Broock, Scheinkman, Dechert, & LeBaron, 1996). A series of conventional linear tests, as those mentioned in the introduction, were run beforehand and are available from the authors upon request. These tests are incapable of capturing nonlinear patterns in the data.

The series were adjusted an $AR(1)$, $AR(5)$ and an $AR(4)$, prior to the application of the tests, in order to remove any linear dependence, where the optimal lag was chosen to minimize the Schwarz's Bayesian Information Criterion. Therefore, we reject the null hypothesis, given the significance level of 5% (see table 3), where the pure white noise is explained solely by the nonlinearity of the historic series.

**Table 3.** Nonlinearity test results

| Test | OIL | USD/MXN | IPC |
|---|---|---|---|
|  | $AR(1)$ | $AR(5)$ | $AR(3)$ |
| McLeod-Li test |  |  |  |
| Lag 5 | 0.000 | 0.000 | 0.000 |
| Lag 15 | 0.000 | 0.000 | 0.000 |
| Lag 20 | 0.000 | 0.000 | 0.000 |
|  |  |  |  |
| Engle LM test |  |  |  |
| Lag 5 | 0.000 | 0.000 | 0.000 |
| Lag 15 | 0.000 | 0.000 | 0.000 |
| Lag 20 | 0.000 | 0.000 | 0.000 |
|  |  |  |  |
| BDS test |  |  |  |
| m = 2, ε = 0.5s | 0.000 | 0.000 | 0.000 |
| m = 3, ε = 1s | 0.000 | 0.000 | 0.000 |
| m = 4, ε = 1.5s | 0.000 | 0.000 | 0.000 |

## 3. Methodology

Brooks and Hinich (1999) claim that the cross-bicorrelation test would permit a researcher to identify any existence of nonlinear dependence between two pairs of variables. The size of the sample series is *N*, with two stationary variables $x(t_k)$ and $y(t_k)$. As we are working with the first lagged differences and small sub-samples of the total series, the assumption of stationarity is reasonable. Each series is separated into equal length non-overlapping moving time windows or frames, where *t* is an integer and *k* represents the *k*-th window, and both series are jointly covariance stationary, which have been standardized. The test's null hypothesis states that the two variables $x(t_k)$ and $y(t_k)$ have no dependence and in fact are pure white noise. The alternative hypothesis states that the series have cross-bicovariances, $C_{xxy}(r,s)$ defined as $E[x(t_k)x(t_k+r)y(t_k+s)]$, different from zero.

Under the null hypothesis, $C_{xxy}(r,s)$ is zero for every $r,s$ except when $r=s=0$. According to the test, there is dependence between a pair of variables if $C_{xxy}(r,s) \neq 0$ for at least one value of $r$ or a pair of values of $r$ and $s$. Next, we present the statistics that give the $r,s$ sample $xxy$ cross-bicorrelation

$$C_{xxy}(r,s) = (N-m)^{-1} \sum_{t=1}^{N-m} x(t_k)x(t_k+r)y(t_k+s)$$

where $m = \max(r,s)$.

We can interpret the cross-bicorrelations, as the degree of relation of the value of one variable, with the value of the cross-correlation of the two variables. The second-order test does not include current elements, and is executed on the errors terms of an $AR(p)$ fit to clean out the univariate autocorrelation arrangement. Thus, current correlations will not be a reason for rejecting the null hypothesis. To perform the third-order test, we apply the test on the error terms of a $VAR(1)$ for USD/MXN-IPC and OIL-IPC, $VAR(2)$ for OIL-USD/MXN; the model having a current term in one of the equation (the order $p$ of the $AR(p)$ and $VAR(p)$ models is chosen to minimize the Schwartz (BIC) criterion). The pre-whitening step is grounded on the elimination of any presence of linear correlation or cross-correlation. Therefore, any outstanding dependence between the variables, should be classified as nonlinear. Let $L = N^c$ where $0 < c < 0.5$ (for our case of study we selected $c = 0.4, N = 4277$, to have 152 non-overlapped windows of length 28 days). The corresponding test statistics for non-zero cross-bicorrelations are,

$$H_{xxy}(N) = \sum_{s=-L}^{L} \sum_{r=1}^{L} (N-m)C_{xxy}^2(r,s), \quad ' - s \neq -1,1,0$$

In these statistics $L$ is the number of times that the correlations are verified and $L(2L-1)$ is the number of times that the cross-bicorrelations are probed. Following Hinich (1996), we state that $H_{xxy}$ are asymptotically $\chi^2$ distributed with $L$ and $L(2L-1)$ degrees of freedom, as $N \to \infty$.

4. **Results and discussion**

In table 4 we present the results of the Brooks and Hinich cross-bicorrelation test. The first column shows the total number of non-overlapped windows, in which the sample was partitioned, the second column the length of each window measured in days, the third column the pair of series for which the cross-bicorrelation is tested, the fourth column shows the number (and percentage from the total) of significative windows, where periods of nonlinearity were found, and the fifth column presents the correlation between the pair of return time series under study.

Table 4. Cross-Bicorrelation test results

| Windows Total | Window | Series | Significative $xxy$ windows | Correlation ($xxy, yyx$) |
|---|---|---|---|---|
| 152 | 28 | OIL-USD/MXN | 73 (48.0 %) | 0.40 |
| 152 | 28 | USD/MXN-IPC | 16 (10.5 %) | 0.65 |
| 152 | 28 | OIL-IPC | 120 (78.9 %) | 0.62 |

As can be observed from the table, there is more cross-bicorrelation or co-movements between Oil and IPC, indicating a clear spillover, although this pair is not the one with the highest correlation from the three pairs of series studied. The higher nonlinear dependence observed in the Oil-IPC might be due to the fact that the Mexican economy is highly dependent on oil exports and when there are high volatility periods in the oil price, the uncertainty in such price might affect the prices of firms listed in the IPC. From Fig. 1, it can be observed that the Oil and IPC time series have the same tendency from which a positive bidirectional relationship can be inferred, although there is a period between 2007 and 2009, where both series decrease and then increase again.

The USD/MXN-IPC pair of return time series is the one with highest correlation (0.65) although with the smallest (10.5%) percentage of cross-bicorrelation windows. We explain such a high correlation from the fact that Mexico, being an oil exporter and a growing emerging economy, attracts foreign investments that come in and out according to financial market expectations, which impacts the exchange rate and the stock market.

Finally, the Oil-USD/MXN pair of return series exhibits 48% of significant cross-bicorrelation windows with the lowest correlation (0.40) of all series under study. The periods of co-movement might be due to the fact that when there is an increase in oil prices it impacts financial markets, appreciating the exchange rate in oil exporting economies, as is the case of Mexico.

In Fig. 2 we show the $(1 − p)$-values of the significant cross-bicorrelation non-overlapped windows, where nonlinear dependence amongst the pairs of return time series occurs. One can observe, how the cross-correlation method allows the researcher to localize in time, the periods of such dependence. All dates of each window are available from the authors upon request.

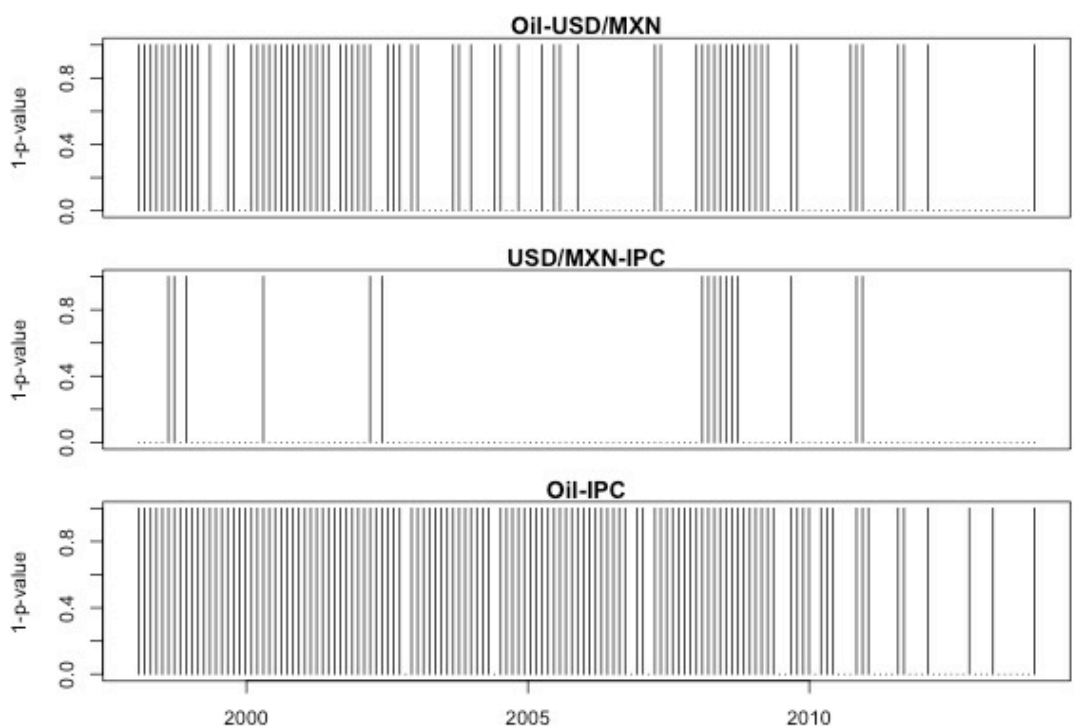

**Fig 2.** ($1-p$)-values of the significant cross-bicorrelation windows by date. Source: The authors.

## 5. Conclusions

We have studied the stationarity and nonlinearity of the return time series of Mexican Crude Oil, USD/MXN exchange rate and Mexican Stock Index (IPC). After proving the series stationary and nonlinear, we used the Brooks and Hinich crosss-bicorrelation statistical test, which enabled the researcher to uncover co-movement, through periods of nonlinear dependence, on pairs of time series. We found a number of significant windows of nonlinear dependence amongst the Oil-IPC (78.9%), Oil-USD/MXN (48%) and USD/MXN-IPC (10.5%) and gave plausible interpretations for such periods of nonlinear dependence. Some of the cross-bicorrelation moments identified, might be prone to a deeper exploration, from an economical point of view and is left as work in progress.

**Acknowledgements**

We thank Rubén Chavarín-Rodriguez for constructive comments that enabled us to improve our manuscript. All possible errors remain the sole responsibility of the authors.